\documentclass   [a4paper,12pt]{article}
\usepackage{graphicx}
\begin{document}

\begin{center}
 {\bf  Schr\"{o}dinger and Hamilton-Jacobi equations   } \\ [5mm]
      Milo\v{s} V. Lokaj\'{\i}\v{c}ek \\
     Institute of Physics, AVCR, 18221 Prague 8, Czech Republic \\
       e-mail: \hspace{5mm}  lokaj@fzu.cz
\end{center}
\vspace{5mm}

Abstract

Time-dependent Schr\"{o}dinger equation represents the basis of
any quantum-theoretical approach. The question concerning its
proper content in comparison to the classical physics has not
been, however, fully answered until now. It will be shown that
there is one-to-one physical correspondence between basic
solutions (represented always by one Hamiltonian eigenfunction
only) and classical ones, as the non-zero quantum potential has
not any physical sense, representing only the "numerical"
difference between Hamilton principal function and the phase of
corresponding wave function in the case of non-inertial motion.
Possible interpretation of superposition solutions will be then
discussed in the light of this fact. And also different
interpretation alternatives of the quantum-mechanical model will
be newly analyzed and new attitude to them will be reasoned.
 \\[5mm]

{\bf  1. Introduction }

Copenhagen quantum mechanics used commonly for the description of
microscopic physical processes is based on Schr\"{o}dinger
equation
\begin{equation}
  i\hbar\frac{\partial}{\partial t}\psi(x,t)=H\psi(x,t), \;\;\;\;
     H=-\frac{\hbar^2}{2m}\triangle + V(x)   \label{schr}
\end{equation}
and on several additional assumptions. Its predictions differ
rather significantly from those of classical physics. And it is
possible to ask how much responsibility for different behavior
lies already in the Schr\"{o}dinger equation and how much in the
other assumptions.

The question was discussed already earlier and it was assumed that
the difference from the classical picture in the Schr\"{o}dinger
equation is given by the existence of the additional quantum
potential in one derived equation being analogous to
Hamilton-Jacobi equation (see, e.g., Ref. \cite{bohm}). Some
authors tried then to explain the existence of this potential as
the consequence of a kind of Brownian motion exhibited by
classical microscopic particles \cite{nels,pena,smol}. Hoyer
showed then that Schr\"{o}dinger equation may be derived when the
classical behavior is combined with the Boltzmann probability
\cite{hoyer}, and similar approach was proposed earlier also by
Ioannidou \cite{ioan}.

We should like to continue in the discussion of the problem in a
somewhat different way. Let us assume that a system consisting of
$N$ particles is to be described by both the considered equations.
To perform more systematic analysis it is then useful to divide
the solutions of Schr\"{o}dinger equation into two different
classes:
\\
 - basic solutions, when the wave function $\psi(x,t)$ exhibits
exponential time dependence and the space part of its is given by
one Hamiltonian eigenfunction;  \\
 - superposition solutions being formed by different linear superpositions
of basic solutions.

The Hamiltonian $H$ represents the total (kinetic + potential)
energy of all particles. And it is possible to define their
positions and momenta (velocities) in the usual way, i.e. as
expectation values of corresponding operators at any $t$. The
solutions of Schr\"{o}dinger equation may be then correlated to
the solutions of Hamilton equations (see Sec. 2). It will be shown
in Sec. 3 that in the case of inertial motion the phase of wave
function is quite identical with Hamilton principal function.
Non-inertial case will be discussed in Sec. 4; one-dimensional
case being analyzed to greater detail. Possible interpretation of
superposition solutions that have not any direct counterpart in
the classical physics will be discussed in Sec. 5. Different
interpretations of quantum-mechanical model in correlation to
superposition solutions will be then dealt with in Sec. 6.
\\

{\bf  2. Relation between Schr\"{o}dinger and Hamilton-Jacobi
equations }

The complex wave function fulfilling Eq. (\ref{schr}) may be
written in the form
\begin{equation}
      \psi(x,t)
      \;=\; \lambda(x,t)\, e^{\frac{i}{\hbar}\Phi(x,t)}  \label{psi}
\end{equation}
and the  Schr\"{o}dinger equation may be substituted  by two
equations
\begin{eqnarray}
 \frac{(\nabla \Phi)^2}{2m}\,+\, V\,+\,V_q
                       &=& -\,\partial_t\,\Phi \;,   \label{hamj}
                        \\
 \triangle\Phi \,+\,2(\nabla\,\Phi)(\nabla\,lg\,\lambda) &=&
       -2m\;\partial_t\, lg\,\lambda   \label{ham2}
\end{eqnarray}
for two real functions: modulus $\lambda(x,t)$ and phase
$\Phi(x,t)$. It holds in general case
\begin{equation}
     V_q(x,t)\,=\,-\frac{\hbar^2}{2m}\frac{\triangle\lambda}{\lambda}
\end{equation}
and Eq. (\ref{hamj}) may be correlated to Hamilton-Jacobi equation
\begin{equation}
      \frac{1}{2m}(\nabla S(x,t))^2 + V(x) = -{\partial_t S(x,t)}   \label{haja}
\end{equation}
where $S(E,t)$ is substituted by $\Phi(x,t)$ and the potential
$V(x)$ by
\begin{equation}
    V_t(x,t)\,=\,V(x)\,+V_q(x,t);     \label{Vt}
\end{equation}
the additional term $V_q(x,t)$ being denoted usually as quantum
potential.
\\

{\bf 3.  Equivalence for inertial motion }

The Hamilton-Jacobi equation contains Hamilton principle function
$S(x,t)$ that determines the momentums of particles in given time:
      \[   p(x,t)=\nabla S(x,t).  \]
 And one should ask whether or when the solutions of Schr\"{o}dinger
equation may describe the same physical systems as those of
Hamilton-Jacobi equation or what is the difference.

We will limit ourselves to time-independent potential $V(x)$. For
the basic solutions of Schr\"{o}dinger equation it is then
possible to write
\begin{eqnarray}
             \psi^{(E)}(x,t) \;&=&\; \psi_E(x) e^{-\frac{i}{\hbar}Et},    \label{eps} \\
              H\psi_E(x) \;&=&\; E\psi_E(x);         \label{eha}
\end{eqnarray}
$E$ is the corresponding energy value, phase
$\Phi^{(E)}(x,t)=-Et+\Phi_E(x)$ and modulus $\lambda^{(E)}(x) =
\lambda_E(x)$ (see Eq. (\ref{psi})) is independent of $t$.

Let us start with the simple case of inertial motion of one
particle, i.e., by putting  $V(x) \equiv 0$. It holds then
     \[  \Phi^{(E)}(x,t) =  - Et + x\sqrt{2mE}, \;\;\;\; \lambda_E(x) = 1; \]
and consequently, $V_t=V_q=0$.  It holds also
           \[ S(x,t) \;=\; \Phi(x,t),  \]
which means that the constant particle momentum equals
         \[ p\,=\, \frac{\partial S(x,t)}{\partial x}\,=\,
                        \frac{\partial\Phi_E(x)}{\partial x} .  \]
\\

 {\bf 4. Basic solutions with separated x- and t-dependencies   }

In a more general case, when  $V(x)\neq 0$, there is a certain
difference between $S(x,t)$ and $\Phi(x,t)$  as the additional
potential term $V_q(x)$ may be non-zero. Let us limit to the basic
solutions represented by Eqs. (\ref{eps}) and (\ref{eha}) and let
us write
 \[   \psi_E(x) \,=\,\lambda_E(x)e^{\frac{i}{\hbar}\Phi_E(x)} . \]

It is then possible to put
   \[ {\bf P}_E(x) \,=\, \nabla\Phi_E(x); \]
for basic solutions it holds also
           \[ {\bf P}_E(x)  \,=\,\nabla\Phi^{(E)}(x,t). \]
 Eq. (\ref{ham2}) may be then rewritten as
\begin{equation}
  \triangle\Phi_E(x) \,+\, 2\,{\bf P}_E(x).\nabla\lg\lambda_E(x)\,=\,0  \label{PE}
\end{equation}
and the function $\lambda_E(x)$ might be determined with the help
of $\Phi_E(x)$ or of ${\bf P}_E(x)$ components.

The last equation might be easily solved in {\it one-dimensional}
case. It should hold then
     \[ \lambda_E(x)\,=\, P_E^{-1/2}   \]
as $\partial^2_x \Phi_E(x)\,=\,\partial_x P_E(x)$. It might lead,
however, in some cases to unacceptable behavior (e.g., in the case
of an oscillating system the quantum potential would be divergent
in some points) which may be avoided in one-dimensional case only
if $\;\Phi_E(x)=const$; Eq. (\ref{ham2}) loosing any sense. And
the function $\lambda(x)$ (or quantum potential) may be
established in such a case on the basis of Eq. (\ref{hamj}) with
the help of the condition
\begin{equation}
   \frac{\partial^2_x\lambda_E(x)}{\lambda_E(x)} =
                   \frac{2m}{\hbar^2}(E-V(x))  .
\end{equation}
E.g., in the case of one-dimensional harmonic oscillator it holds
for individual quantum numbers $n$:
\begin{equation}
   \frac{\partial^2_x\lambda_n(x)}{\lambda_n(x)} =
      -\frac{2m}{\hbar^2}[(n+\frac{1}{2})\hbar\omega-\frac{m\omega^2}{2}x^2] ,
\end{equation}
which may be verified when $\lambda_n(x)$ is derived from the
eigenfunctions of corresponding Hamiltonian.

The solution in a more-dimensional case represents, of course,
more complicated problem. One may expect, however, similar results
to be obtained.

The quantum potential goes always to zero if the motion of all
involved objects is inertial, i.e., if any forces do not act
between them. It is non-zero in the case of non-inertial motion,
depending on corresponding characteristics of individual physical
 states.
It represents the diference between Hamilton principal function
$S(x,t)$ (fulfilling Eq. (\ref{haja})) and the phase $\Phi(x,t)$
of wave function (fulfilling Eq. (\ref{hamj})) that are expressed
by identical functions in the case of inertial motion.

The quantity $\nabla\Phi_E(x,t)$ fulfilling the condition
\begin{equation}
    (\nabla \Phi_E(x))^2 \,=\, 2m\,(E\,-\,V_t(x))       \label{SE}
\end{equation}
does not represent the momentum components of corresponding
particle in non-inertial case; $V_t(x)$ being defined by Eq.
(\ref{Vt}). And the quantum potential has not practically any
actual physical sense.
\\

{\bf 5. Superpositions of basic solutions   }

It has been shown in Sec. 4 that the basic solutions of
Schr\"{o}dinger equation may be correlated to corresponding
solutions of Hamilton-Jacobi equation; both representing the same
physical behavior. There is, however, a difference; any
superposition of basic solutions is again a solution of
Schr\"{o}dinger equation, which has not its analogue as to
Hamilton-Jacobi equation. It follows from the fact that the
Hamilton-Jacobi equation is non-linear differential equation,
while the Schr\"{o}dinger equation is linear differential
equation.

Let us assume now that the given set of solutions belongs to
continuous energy spectrum. It means that any superposition of
basic states must correspond to energy $E$ lying in the same
energy spectrum, but representing a different state than the
corresponding basic solution. It cannot be characterized by any
Hamilton principal function and described as a direct solution of
Hamilton-Jacobi equation.
 And one should ask what is the difference between the basic
solutions (corresponding to classically moving particles) and
their superpositions, and how to interpret these superpositions
physically.

Individual superpositions may hardly represent simple physical
states with given energy values. The superpositions of different
basic states of Schr\"{o}dinger equation may provide, however, an
advantage in describing measurement results concerning microscopic
objects. Any measurement of such a type is based in principle on
studying mutual collisions between microscopic particles or
interactions of these particles with a macroscopic device. In both
the cases the exact classical description would require to know
the value of impact parameter between mutually colliding
microscopic particles or between a microscopic particle and a
collision center in the microscopic structure of a macroscopic
object (i.e., the value of the corresponding "hidden" parameter in
the quantum-mechanical picture).

However, the impact parameter is always statistically distributed,
even if the other physical properties are the same; e.g., the
initial kinetic energy (or momenta) of mutually colliding objects.
One cannot determine exactly mutual initial positions of colliding
particles in the direction perpendicular to particle tracks.
 Consequently, it is also the total energy (i.e., the sum of
kinetic and potential energies) that is statistically distributed.
The dependence on the value $b$ of impact parameter in individual
events is given in principle by the shape of mutual potential
function $V(b)$. Statistical distributions of other physical
characteristics may, of course, contribute to the distribution of
the total energy value $E$, too.

To describe such a situation with the help of Hamilton-Jacobi
equation must be denoted, of course, as practically impossible. It
would be necessary to sum (or to integrate) individual solutions
in agreement with a statistical weighting function. The final
result may be obtained, of course, directly by solving
Schr\"{o}dinger equation with the initial function $\psi_0(x)$
given by a corresponding statistical superposition of basic
states:
\begin{equation}
          \psi_0(x) \,=\, \int\!dE\, c(E)\, \psi_E(x), \;\;
                    \int\!dE\, c^*(E)\,c(E) = 1      \label{sup}
\end{equation}
where $\psi_E(x)$ fulfills Eq. (\ref{eha}) and the weighting
function $c(E)$ characterizing the corresponding superposition may
be real.
 One general solution of Schr\"{o}dinger equation may represent,
therefore, the statistically distributed  result of the whole
measurement process concerning microscopic objects.

The collision processes in considered experiments (i.e., mutually
colliding free particles at a given energy E) may be described by
eigenstates belonging to continuous Hamiltonian spectrum. In the
case of bound states the eigenvalue spectrum is discrete, but
individual basic states correspond to the solutions of
Hamilton-Jacobi equation, forming a subset of classically admitted
continuous set of solutions.

And it is the Schr\v{o}dinger equation that corresponds better
than Hamilton-Jacobi equation to experimental data in the
microscopic region. Consequently, on the microscopic level the
Schr\"{o}dinger equation should be preferred to classical
description. However, the Schr\v{o}dinger equation may be applied
in principle also to physical systems with high energy values
(including classical systems) without assuming the Planck constant
(being of general validity) to change its value. The differences
between the admitted energy values will be so small that it will
not be possible to register them at all.
\\

{\bf  6.  Superposition solutions and different interpretations
          of quantum-mechanical mathematical model}

As to the quantum mechanics there are two different
interpretations of solutions of Schr\"{o}dinger equation being
discussed in literature all the time: orthodox (Copenhagen) or
statistical (ensemble); see, e.g., \cite{home}.
 In the latter case one may assume, that a superposition
solution describes measurement process when an amount of
identically prepared microscopic systems is measured, each of them
being characterized by different values of some randomly
distributed hidden parameters (e.g., impact parameter as discussed
in Sec. 5). The corresponding result may be, therefore, easily
obtained with the help of Schr\"{o}dinger equation if the initial
state is expressed as corresponding superposition of basic states
(see Eq. (\ref{sup})).

As to the former interpretation two other important assumptions
have been added to the basic Schr\"{o}dinger equation: \\
 - all solutions of its have been assumed to be represented by
vectors in the standard Hilbert space that has been spanned on one
set of Hamiltonian eigenfunctions; \\
 - the superposition principle (holding in any mathematical metric
space) has been interpreted as the physical property of a
described physical system.

It follows from these additional assumptions that any vector of
the given Hilbert space must represent a physical state, which
leads to an important contradiction concerning experimental data
for bound states. As already mentioned in Sec. 5 it follows from
Schr\"{o}dinger equation that the corresponding basic states form
a set of discrete states only. However, according to additional
Copenhagen assumptions all superpositions (exhibiting continuous
spectrum in the whole corresponding energy interval) should exist,
too, which contradicts experimental facts.

Some discrepancy between the Schr\"{o}dinger equation and the
additional assumptions of Copenhagen model was indicated by Pauli
\cite{pauli} already in 1933. He showed that the time dependence
following from Schr\"{o}dinger equation may be represented fully
in the given standard Hilbert space only if the Hamiltonian
exhibits the spectrum of all real values from -$\infty$ to
+$\infty$. Many attempts of removing this deficiency have been
done in the past, but they have not been successful.

It has been shown only recently (see \cite{kun1,kun2}) that the
criticism of Pauli may be removed if the standardly used Hilbert
space is extended (doubled). It should consist from two mutually
orthogonal simple Hilbert spaces (discussed already earlier by Lax
and Phillips \cite{lax1,lax2}), which, e.g., in the case of two
free particles may be interpreted as subspaces corresponding to
incoming or outgoing states.

Then, of course, all classically required parameters should be
included in the description of any physical system in an
equivalent way  as other characteristics; e.g., also the impact
parameter in the case of two particles even if actual measurement
may concern the set of systems exhibiting statistically
distributed values. The latter interpretation of Schr\"{o}dinger
equation is then in full harmony with the model based on the
extended Hilbert space and all superpositions of basic states
should be interpreted as considered in Sec. 5.
\\

{\bf  7.  Conclusion  }

Schr\"{o}dinger's discovery has consisted in that it has been
possible to describe simple classical systems alternatively with
the help of the wave equation. Schr\"{o}dinger demonstrated it
practically in the case of inertial motion. It has been shown in
the preceding that it holds for any basic states, i.e., for all
solutions of Schr\"{o}dinger equation characterized by one
eigenfunction of Hamiltonian only.

The general solutions of Schr\"{o}dinger equation (i.e.,
superpositions of basic solutions) may be then made use of in
describing statistically distributed results of measurement
processes concerning microscopic objects. One superposition
solution may represent the statistical result of a whole
measurement process if the weighting function is correspondingly
chosen. That is in full agreement with the results of Hoyer
\cite{hoyer} and Ioannidou \cite{ioan} who have shown that
Schr\"{o}dinger equation may be derived if the classical behavior
is combined with a kind of statistical distribution.

And one must ask if the time has not come when it is necessary to
put the question whether the statistical (ensemble) interpretation
of quantum mechanics (i.e., the Schr\"{o}dinger equation alone)
should not be preferred generally and earlier additional
assumptions involved in the orthodox (Copenhagen) quantum
mechanics abandoned.

The difference in the mathematical model in such a case would
consist in the extension (doubling) of the standardly used Hilbert
space as shown in \cite{kun1,kun2}. And it is also the
mathematical superposition principle that may be hardly
interpreted in a physical sense. Otherwise, nothing else would
change in the earlier approach based on Schr\"{o}dinger equation,
including the existence of discrete bound states and other quantum
characteristics.
\\


{\footnotesize
\begin{thebibliography}{99}
\bibitem{bohm}
D. Bohm: A suggested interpretation of the quantum theory in terms
of "hidden variables"; {\it Phys. Rev.}, Vol. 85, (1952), pp.
180-93.
\bibitem{nels}
E. Nelson: Derivation  of the Schr\"{o}dinger equation from
Newtonian Mechanics; {\it Phys. Rev.}, Vol. 150, (1969), pp.
1079-85.
\bibitem{pena}
L.de la Pe$\tilde{n}$a-Auerbach: New formulation of stochastic
theory and quantum mechanics; {\it Lett. Math. Phys.}, Vol. 10
(1969), 1620-9.
\bibitem{smol}
L. Smolin: Matrix models as non-local hidden variables theories;
{\it "Quo vadis quantum mechanics?"} (eds. A.Elitzur et al.),
Springer, 2005, pp. 121-52.
\bibitem{hoyer}
U. Hoyer: {\it Synthetische Quantentheorie}; Georg Olms Verlag,
Hildesheim, 2002.
\bibitem{ioan}
H. Ioannidou: A new derivation of Schr\"{o}dinger equation; {\it
Lett. al Nuovo Cim.}, Vol. 34, (1982), pp. 453-8.
\bibitem{home}
D.Home, M.A.B.Whittaker: Ensemble interpretations of quantum
mechanics. A modern perspective; {\it Phys. Rep. C}, Vol.
21(1992), 223-317.
 \bibitem{pauli}
W.Pauli: Die allgemeinen Prinzipien der Wellenmechanik; {\it
Handbuch der Physik XXIV}, Springer, Berlin 1933, p. 140.
\bibitem{kun1}
P.Kundr\'{a}t, M.Lokaj\'{\i}\v{c}ek: Three-dimensional harmonic
oscillator and time evolution in quantum mechanics; {\it Phys.
Rev. A}, Vol. 67, (2003), art. 012104.
\bibitem{kun2}
P.Kundr\'{a}t, M.Lokaj\'{\i}\v{c}ek: Irreversible time flow and
Hilbert space structure; {\it New Research in Quantum Physics}
(eds. Vl.Krasnoholovets, F.Columbus), Nova Science Publishers,
Inc., 2004, pp. 17-41.
\bibitem{lax1}
P.D.Lax, R.S.Phillips: {\it Scattering theory}; Academic Press,
New York, 1967.
\bibitem{lax2}
P.D.Lax, R.S.Phillips, {\it Scattering theory for automorphic
functions} (Princeton 1976)
 \end{thebibliography}  }
\end{document}